\documentclass[runningheads]{llncs}
\usepackage{float}
\usepackage{graphicx}
\usepackage{listings}
\usepackage{xcolor}
\usepackage{hyperref}
\usepackage{listings}

\begin{document}

\title{Comparing Llama-2 and GPT-3 LLMs for HPC kernels generation}
\titlerunning{}

\author{
Pedro Valero-Lara\inst{1}$^,$$^*$\orcidID{0000-0002-1479-4310}, \\
Alexis Huante\inst{2}\orcidID{0009-0008-2818-0265}\\
Mustafa Al Lail\inst{2}\orcidID{0009-0000-0326-6363}, \\
William F. Godoy\inst{1}\orcidID{0000-0002-2590-5178}, \\
Keita Teranishi\inst{1}\orcidID{0000-0001-6647-2690}, \\
Prasanna Balaprakash\inst{1}\orcidID{0000-0002-0292-5715}, \\
Jeffrey S. Vetter\inst{1}\orcidID{0000-0002-2449-6720}
}
\authorrunning{Valero-Lara et al.}

\institute{Oak Ridge National Laboratory, Oak Ridge, TN, 37830, USA \and 
Texas A\&M International University, Laredo, Texas 78041, USA \\
$^*$Corresponding author: \email{valerolarap@ornl.gov}
}

\maketitle              
\begin{abstract}
We evaluate the use of the open-source Llama-2 model for generating well-known, high-performance computing kernels (e.g., AXPY, GEMV, GEMM) on different parallel programming models and languages (e.g., C\texttt{++}: OpenMP, OpenMP Offload, OpenACC, CUDA, HIP; Fortran: OpenMP, OpenMP Offload, OpenACC; Python: numpy, Numba, pyCUDA, cuPy; and Julia: Threads, CUDA.jl, AMDGPU.jl). We built upon our previous work that is based on the OpenAI Codex, which is a descendant of GPT-3, to generate similar kernels with simple prompts via GitHub Copilot. Our goal is to compare the accuracy of Llama-2 and our original GPT-3 baseline by using a similar metric. Llama-2 has a simplified model that shows competitive or even superior accuracy. 
We also report on the differences between these foundational large language models as generative AI continues to redefine human-computer interactions.
Overall, Copilot generates codes that are more reliable but less optimized, whereas codes generated by Llama-2 are less reliable but more optimized when correct.
\end{abstract}


\section{Introduction}
\label{sec:Introduction}

Generative-AI large language models (LLMs) are transforming the software industry by automating manual tasks, such as developing, testing, and deploying applications. The use of LLMs could lead to faster and more cost-effective software development. LLMs are also revolutionizing entertainment, education, and healthcare industries by creating realistic images, text, music, and code. However, there are social and ethical concerns surrounding LLMs, including the risk of deep fakes being created and distributed as misinformation or to harm individuals. Therefore, the risks and benefits of LLMs must be carefully considered before widespread adoption. 

The emergence of exascale computing presents a challenge in developing software for high-performance computing (HPC) systems owing to the varying hardware and programming models in these complex architectures. To address this challenge, AI-assisted code generation could be used. LLMs can generate code in commonly used programming languages, including C\texttt{++}, Fortran, Python, and Julia. This innovation could make software development for HPC more efficient and manageable. However, limitations exist for AI-assisted code generation given it may only sometimes produce code that is as efficient or reliable as human-written code. The current state of practice, the limitations, and the potential of LLMs must be fully understood to realize their benefits.

The effort described in this paper builds on our previous work~\cite{abs-2306-15121}, in which we investigated the effectiveness of OpenAI Codex for generating HPC code for various numerical kernels in different programming languages and models, including C\texttt{++}, Fortran, Python, and Julia. The study found that the output of OpenAI Codex for C++ is closely linked to the popularity and sophistication of programming models. For example, OpenMP~\cite{openmp}  and CUDA~\cite{cuda} received high scores because they are widely used and well-established programming models. However, HIP~\cite{hip} received lower scores because it is a newer programming model that is not as widely used. The study also found that prompts in Fortran or Python can benefit from incorporating code keywords. However, Julia's prompts perform adequately without the need for code keywords for its mature HPC programming models.

This paper also describes our evaluation of Meta AI's LLM (Llama-2) for generating HPC kernels. The version of Llama-2 we used has 70 billion parameters and was provided by Hugging Chat, an open-source chat bot platform that relies on LLMs to power its conversations. This platform is built on top of the Hugging Face ecosystem. Our evaluation involves generating code for three fundamental numerical kernels: AXPY, GEMV, and GEMM. We then test the resulting 144 kernel codes in four programming languages,
 C++, Fortran, Python, and Julia, by using various programming models and compilers. These included OpenMP, OpenACC~\cite{openacc}, CUDA, HIP, numpy~\cite{van2011numpy}, Numba~\cite{lam2015numba}, cuPy~\cite{nishino2017cupy}, pyCUDA~\cite{KLOCKNER2012157}, Julia's Base Threads~\cite{knopp2014experimental}, CUDA.jl~\cite{besard2018juliagpu}, and AMDGPU.jl~\cite{AMDGPU}.

The paper is organized as follows: Section~\ref{sec:Related Work} provides an overview of related efforts that have brought attention to these topics in computer science. Section~\ref{sec:Methodology} outlines our methodology for generating and evaluating the code with Llama-2. In Section~\ref{sec:Results}, we present the results of our evaluation and our findings for each language, kernel, and programming model along with additional keyword inputs on the generated outputs. Finally, Section~\ref{sec:Conclusions} presents our conclusions.

\section{Related Work}
\label{sec:Related Work}

The Generative Pre-trained Transformer 3 (GPT-3)~\cite{NEURIPS2020_1457c0d6}
is a game changer in the evolution of human-computer interactions. Developed by OpenAI,\footnote{\url{https://openai.com/}} GPT-3 is the third generation of the prediction-based foundational LLM used for several AI-generated, human-like text applications. GPT-3 is used in several natural language processing tasks~\cite{doi:10.1126/science.aaa8685}, including ChatGPT, due in part to the large investment (\$12 million USD) and size of its training model (175 billion parameters at 800\,GB). Hence, GPT-3 and its successor GPT-4\footnote{\url{https://openai.com/product/gpt-4}} are defining several societal questions for the near future.
Today, we are at the beginning of a race to develop the best LLM model. In addition to GPT, we can find recently released foundational LLMs such as Llama-2~\cite{abs-2307-09288} and PaLM 2\footnote{\url{https://ai.google/discover/palm2/}}.

As we enter the exascale computing era, which is dominated by the extreme heterogeneity of hardware and programming models~\cite{vetter:2018:extreme}, AI-assisted code generation could play a key role in how we develop, deploy, and test software that targets HPC systems. Traditional human-readable code in languages such as C\texttt{++}~\cite{stroustrup2013c++}, Fortran~\cite{Fortran}, Python~\cite{Python}, and more recently Julia~\cite{Bezanson2017-ca}, are a straightforward application for LLM's capabilities---capabilities that could help redefine software development. In fact, this rapidly evolving field was recently surveyed in our previous work~\cite{abs-2306-15121}, in which we evaluated the performance of the GPT-3 descendant OpenAI Codex for HPC kernel generation by using GitHub Copilot for several parallel programming models. The quality of the responses depends largely on the number of repositories and programming model maturity.
Nichols et al.~\cite{nichols2023modeling} fine-tuned the use of LLMs to improve the generation of OpenMP pragmas in parallel algorithm implementations, including MPI cases. Chen et al.~\cite{chen2023lm4hpc} presented LM4HPC, a framework to conduct HPC-specific tasks in the context of LLMs, and highlighted the lack of training and evaluation datasets in HPC. Hence, we expect to see more work in the convergence of HPC and generative AI via LLMs because of the field's rapid evolution. To the best of our knowledge, this is the first evaluation of Llama-2 for the generation and correctness of HPC kernels and comparison to our baseline from previous work.
\section{Methodology}
\label{sec:Methodology}

First, we use prompts similar to those in our previous research~\cite{abs-2306-15121}, which are simple prompts based on the programming language, kernel, and programming model. The quality of the prompt is important because it determines how the LLM will generate the requested code based on the information provided. So, several adjustments were made to the prompt until Llama-2 was outputting the code requested. Importantly, the output from the LLM also depends on the data used to train the model. For example, the LLM may not be trained well enough for a particular language or model and may therefore produce inaccurate code no matter the prompt given. 

The methodology used in this study involves two main characteristics that will be discussed in the next subsections:
    (1) how we prompted Llama-2 for code generation based on the kernel, parallel programming model, and programming language and
    (2) a code correctness metric that will be evaluated by expert observation.

\subsection{Experimental Setup}
\begin{figure}[h]
    \centering
    \includegraphics[width=0.75\textwidth,height=0.33\textheight]{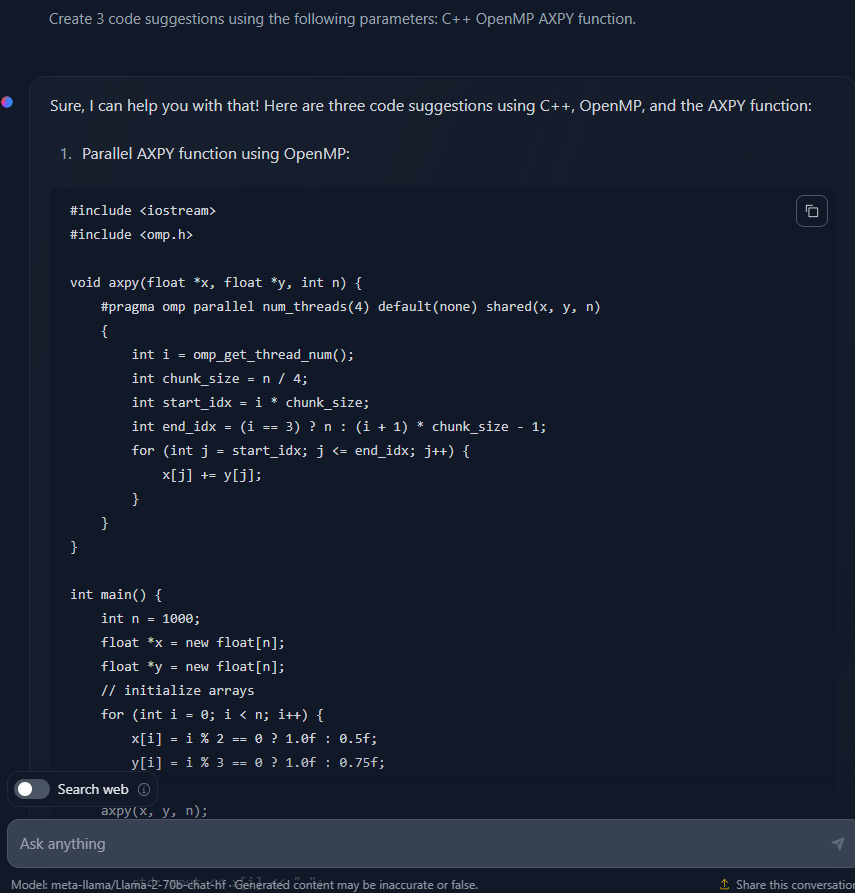}
    \caption{Hugging Chat website interface.}
    \label{fig:HuggingChat-2}
\end{figure}

For our experiments, we used the \href{https://huggingface.co/chat/}{Hugging Chat} website, which, as of August 2023, uses the largest model of Llama-2 called Llama-2-70B. We created an account on Hugging Face to access the necessary features. As shown in Figure~\ref{fig:HuggingChat-2}, the website features a chat box for the user to input their query for the LLM. 

An example of the prompt and the generated code on Llama-2 is illustrated in Figure~\ref{fig:HuggingChat-2}. The structure of the prompt is as follows:

\begin{itemize}
    \item Create 3 code suggestions using the following parameters: 〈Programming Language〉 〈Programming Model〉 〈Kernel〉 〈Keyword〉.
    \item Create 3 code suggestions using the following parameters: 〈Programming Language〉 〈Programming Model〉 〈Kernel〉.
\end{itemize}

Unlike our previous study based on the GitHub Copilot model~\cite{abs-2306-15121}, which can provide one or more codes, we must specify the number of code suggestions we want when using Llama-2. Importantly, the first prompt is used for C\texttt{++}, Fortran, and Python, whereas the second prompt is used only for Julia. This is because, according to previous research, they determined there was slight sensitivity in Julia prompts when adding a keyword~\cite{abs-2306-15121}. For the 〈Kernel〉 section, instead of prompting ``GEMV'' or ``GEMM,'' we used the full form of the abbreviations, which are ``general matrix-vector multiply'' and ``general matrix-matrix multiply,'' respectively. This is because Llama-2 does not interpret what the abbreviations mean. Additionally, Llama-2 has a character limit, so when prompting for three code suggestions, sometimes it could not finish all three codes. Whenever this was the case, we prompted the LLM to continue with the code generation by saying, ``please continue with the code,'' ``you stopped, please continue,'' or similar.

Next, Table~\ref{table:1scope} lists all the programming languages, programming models, and keywords used in this study. We used the AXPY, GEMV, and GEMM kernels for every programming model.
These kernels correspond to one specific operation of the three different levels of the Basic Linear Algebra Subprograms (BLAS) library:\footnote{\url{https://www.netlib.org/blas/}}
the AXPY level-1 BLAS routine computes a scalar-vector multiplication, the GEMV level-2 BLAS routine computes a matrix-vector multiplication, and the GEMM  level-3 BLAS routine computes a matrix-matrix operation. The BLAS library operations increase in complexity with each level. Also, the higher the level of the BLAS routine, the more possibilities for optimizations.   

We used a total of 48 prompts, which resulted in 144 codes generated by Llama-2. These codes will be evaluated by the correctness metric described in the next subsection, and we will compare the results to those of the LLM Copilot model from earlier work~\cite{abs-2306-15121}.

\begin{table}[]
\centering
\begin{tabular}{lll}
\hline
\multicolumn{3}{c}{Kernels: AXPY, GEMV, GEMM}                                                                          \\ \hline
\multicolumn{1}{|l|}{Programming Language} & \multicolumn{1}{l|}{Programming Model}  & \multicolumn{1}{l|}{Keyword}    \\ \hline
\multicolumn{1}{|l|}{C++}                  & \multicolumn{1}{l|}{OpenMP}             & \multicolumn{1}{l|}{function}   \\
\multicolumn{1}{|l|}{}                     & \multicolumn{1}{l|}{OpenMP(offload)}    & \multicolumn{1}{l|}{function}   \\
\multicolumn{1}{|l|}{}                     & \multicolumn{1}{l|}{OpenACC}            & \multicolumn{1}{l|}{function}   \\
\multicolumn{1}{|l|}{}                     & \multicolumn{1}{l|}{CUDA}               & \multicolumn{1}{l|}{function}   \\
\multicolumn{1}{|l|}{}                     & \multicolumn{1}{l|}{HIP}                & \multicolumn{1}{l|}{function}   \\ \hline
\multicolumn{1}{|l|}{Fortran}              & \multicolumn{1}{l|}{OpenMP}             & \multicolumn{1}{l|}{subroutine} \\
\multicolumn{1}{|l|}{}                     & \multicolumn{1}{l|}{OpenMP(offload)}    & \multicolumn{1}{l|}{subroutine} \\
\multicolumn{1}{|l|}{}                     & \multicolumn{1}{l|}{OpenACC}            & \multicolumn{1}{l|}{subroutine} \\ \hline
\multicolumn{1}{|l|}{Python}               & \multicolumn{1}{l|}{numpy}              & \multicolumn{1}{l|}{def}        \\
\multicolumn{1}{|l|}{}                     & \multicolumn{1}{l|}{Numba}              & \multicolumn{1}{l|}{def}        \\
\multicolumn{1}{|l|}{}                     & \multicolumn{1}{l|}{pyCUDA}             & \multicolumn{1}{l|}{def}        \\
\multicolumn{1}{|l|}{}                     & \multicolumn{1}{l|}{cuPy}               & \multicolumn{1}{l|}{def}        \\ \hline
\multicolumn{1}{|l|}{Julia}                & \multicolumn{1}{l|}{Threads}            & \multicolumn{1}{l|}{}           \\
\multicolumn{1}{|l|}{}                     & \multicolumn{1}{l|}{CUDA}               & \multicolumn{1}{l|}{}           \\
\multicolumn{1}{|l|}{}                     & \multicolumn{1}{l|}{AMDGPU}             & \multicolumn{1}{l|}{}           \\
\hline
\end{tabular}
\caption{Parameters used for code generation}
\label{table:1scope}
\end{table}

\subsection{Correctness metric}
To evaluate the correctness of the generated codes, we use the simple metric approach from our previous work~\cite{abs-2306-15121}.
We consider five levels of correctness and proficiency labels between [0], or {\it non-knowledge},  and [1], or {\it expert}, when observing the suggested answers provided by Llama-2 for each combination in Table~\ref{table:1scope}.

\begin{itemize}
    \item [0] {\it non-knowledge}: No code at all or not a single correct code.
    \item [0.25] {\it novice}: One correct code, but it includes other several correct or incorrect programming models (e.g., OpenACC suggestions in an OpenMP prompt).
    \item [0.5] {\it learner}: One correct code, and there are other incorrect codes, but all of them use the requested programming model.
    \item [0.75] {\it proficient}: All codes are correct and use the programming model requested.
    \item [1] {\it expert}: Only one piece of code is provided, and it is totally correct.
\end{itemize}

As mentioned, to make the analysis similar to our previous study on the GitHub Copilot LLM, and to obtain more than one code from Llama-2, we must specify the number of codes that we want. So, we will use the highest metric (expert) for cases in which Llama-2 generates all the three requested codes and does so correctly.

\section{Results}
\label{sec:Results}
The following subsections describe our evaluation of the HPC kernels generated by the Llama-2 LLM for 
four different programming languages: C++, 
Fortran, Julia, and Python. The code generated by Llama-2 has also been collected and uploaded to a GitHub repository.\footnote{\url{https://github.com/mustafalail/Llama-2-70b-HPC-Kernels}} 

\subsection{C\texttt{++}}

C\texttt{++} has become the primary programming language used for heterogeneous HPC architectures due to the support that the open-source and vendor communities provide in terms of programming models and compilers.
Examples include OpenMP, OpenACC, and CUDA, among others such as HIP, Kokkos, and SYCL. In this study, we focused on the most popular, mature, and widely used programming models in the HPC community: OpenMP, OpenACC, CUDA, and HIP.

\subsubsection{OpenMP}
OpenMP is considered the de facto standard for parallel programming. The OpenMP codes generated by Llama-2 have the highest quality among the C\texttt{++} codes. Notably, Llama-2 can leverage relatively advanced OpenMP techniques, including tasking (\texttt{\#pragma omp task}), atomic operations (\texttt{\#pragma omp atomic update}), and single instruction multiple data (SIMD) primitives (\texttt{\#pragma omp simd}), among others (\texttt{\#pragma omp critical}). However, not all codes are correct. Also, in some cases, the OpenMP code provided used a defined number of threads. This is very dependent on the architecture to be used. In general, the number of threads should be equal to the number of cores (\texttt{\#pragma omp parallel num\_threads(4)}). In some particular cases, in the codes corresponding to the AXPY kernel, Llama-2 provided codes that, although similar to the operation conducted by this BLAS routine, were not exactly the same. For instance, the codes did not use a scalar, or they computed other operations, such as dot product. This is not the same for the other operations evaluated (i.e., matrix-vector and matrix-matrix multiplication) in which the codes provided were correct and functional.

We also see significant errors for the OpenMP target offloading case. In most cases, the code generated was a mix of CUDA and OpenMP codes. Also, the OpenMP primitives used did not correspond to OpenMP target offloading. Unlike the previous case, all generated codes were incorrect.

\subsubsection{OpenACC}
A similar scenario is observed in the OpenACC case for the AXPY operation. All codes provided were incorrect and were a mix of CUDA codes with OpenACC primitives. However, much higher quality was found in the other two kernels, in which the OpenACC primitives were effectively used. Indeed, we see some advanced techniques, such as the use of ``collapse'' to enroll two nested and independent for loops (\texttt{\#pragma acc loop independent collapse(2)}). Also, we see an effective movement of data between CPU and GPU in some codes and an effective use of tiling/blocking to decompose the matrices. In this case, the codes provided for the kernels of the matrix-vector and matrix-matrix multiplications were correct. 

\subsubsection{HIP}
For HIP codes, we found the same error in most of the codes that correspond to the computation of the thread index (\texttt{int ind = hipBlockDim\_x * hipBlockIdx\_x + hipThreadIdx\_x;}). In some cases, this index was not even computed or it was only partially computed. This relatively simple error breaks the entire code, even if the rest of the code is correct. Other common errors found include using the same names for both CPU and GPU memory pointers, using bi-dimensional blocks of threads to launch the kernels when the kernel implementation only uses uni-dimensional blocks of threads (or vice-versa), and the wrong use of GPU shared memory. Also, as in the previous OpenMP and OpenACC analyses, we saw a mix of HIP and CUDA codes. In this case, we found errors in all three test cases (AXPY, matrix-vector, and matrix-matrix multiplications).

\subsubsection{CUDA}
Although we found better quality codes for CUDA than for HIP, the Llama-2-generated CUDA codes still contained some important errors. For instance, using \texttt{\_\_device\_\_} function decorators for the kernels implementation when the correct decorator is \texttt{\_\_global\_\_}, wrong name of CUDA library functions (\texttt{hipCublasSdot}), and initializing GPU memory arrays from the CPU are just a few examples of the errors found. However, all of these errors were found in the AXPY kernel. The code generated for the other two kernels was correct and free of errors.
In fact, we observed the effective use of important optimization techniques, such as shared memory (\texttt{\_\_shared\_\_ float smem[32][32];}) and registers (\texttt{register float rA[32];}), which are used to implement relatively complex algorithms based on tiling/blocking for matrix computation.

\subsection{Fortran}
Fortran was one of the first widely used programming languages for HPC back in the 1970s. 
In fact, with reasonably good support for current HPC standards, Fortan is still an important programming language for HPC. In the Fortran community, there are two predominant parallel programming models: (1)~OpenMP, which is more focused on providing parallel codes for CPUs, and (2)~OpenACC, which is more focused on GPUs.

\subsubsection{OpenMP}
Unlike the C\texttt{++} codes generated by Llama-2 for OpenMP, we see much better results from Llama-2 when generating Fortran code for OpenMP, especially for the AXPY routine. All generated codes were correct and made use of a scalar. Also, the code generated for the other two kernels used the OpenMP decorators efficiently. Notably, although no advanced OpenMP primitives (e.g., SIMD, collapse) were used, relatively highly optimized algorithms based on tiling/blocking for the matrix-matrix multiplication kernels were used. Unfortunately, this was not the case for OpenMP target offloading, a case in which all the codes provided did not make correct use of the OpenMP primitives.

\subsubsection{OpenACC}
For OpenACC, Llama-2 provided the wrong OpenACC codes for AXPY kernels and used OpenMP decorators instead of OpenACC ones. Better codes were generated for the other two kernels, and at least one functional code was provided. The OpenACC primitives were not used correctly in many cases, and some of the primitives used do not even exist in the OpenACC standard.

\subsection{Julia}
Julia provides a dynamic, compiled front end to LLVM to target scientific computing and data science. Julia's use in HPC is still an area of active exploration~\cite{GodoyVDTJSMTVC23} and community engagement.
In this section, we evaluate the correctness of three different Julia packages: Base.Threads.jl, CUDA.jl, and AMDGPU.jl, which are used for parallel programming on CPUs, NVIDIA GPUs, and AMD GPUs, respectively.

\subsubsection{Base.Threads.jl}
For the parallel CPU codes that use the Base.Threads.jl Julia package, we found that at least one code provided correct matrix-vector and matrix-matrix multiplication. Unfortunately, this is not the case for the AXPY kernel, and all the codes provided for AXPY were invalid. This could be because of Julia's novelty as a programming language in HPC. Notably, in some cases, it can be challenging to generate different codes that implement exactly the same requested operation, such as AXPY using Base.Threads.jl. Common errors found here include missing keywords (\texttt{@threads}) or the use of other packages (\texttt{Distributed.jl}).

\subsubsection{CUDA.jl and AMDGPU.jl}
The codes generated using the CUDA package (CUDA.jl) were incorrect. Notably, the generated codes attempted to decorate the nested loops that correspond to the kernels in a way that is similar to how they are decorated when using the Base.Threads.jl package.
However, using CUDA.jl is not much different from classic CUDA (i.e., the kernels must be implemented out of the main code, and these must be called/launched by using a very specific syntax [\texttt{CUDA.@sync @cuda threads = threads blocks = blocks kernel(x...)}]).
We found exactly the same issues for the Llama-2-generated AMDGPU.jl codes. 

\subsection{Python}
Python is a high-level, interpreted, general-purpose programming language. 
The Python community is one of the biggest software communities today together with C and C++.\footnote{\url{https://www.tiobe.com/tiobe-index/}}
In this study, we used the most popular parallel solutions in the Python ecosystem: numpy, cuPy, pyCUDA, and Numba.
Like with C++, the codes generated by Llama-2 for the AXPY kernels were incorrect, and they did not compute the AXPY operation.
And again, unlike the AXPY case, Llama-2 provided much better codes for matrix-vector and matrix-matrix multiplication kernels
when using numpy and Numba in particular.

Notably, the quality of these successful cases lies in the use of optimization techniques, such as the decomposition of the matrices into 
chunks or doing stridden memory accesses. However, we found an error that is common in all of the codes generated for cuPy: using the \texttt{\_\_shared\_\_} decorator for the GPU functions instead of the
\texttt{\_\_device\_\_} decorator, which is the one that must be used. Unfortunately, although the rest of the code is correct, this relatively small error breaks the entire code.

\subsection{Llama-2 versus Copilot}

This section compares the results of the GitHub Copilot model against the results presented above.
For the Copilot model, we use the results presented by W. Godoy et al.~\cite{abs-2306-15121}.
The codes generated by the Copilot model are hosted in a GitHub repository.\footnote{\url{https://github.com/keitaTN/Copilot-hpc-kernels}}

\begin{figure}[htb]
\centering
\includegraphics[width=0.48\textwidth]{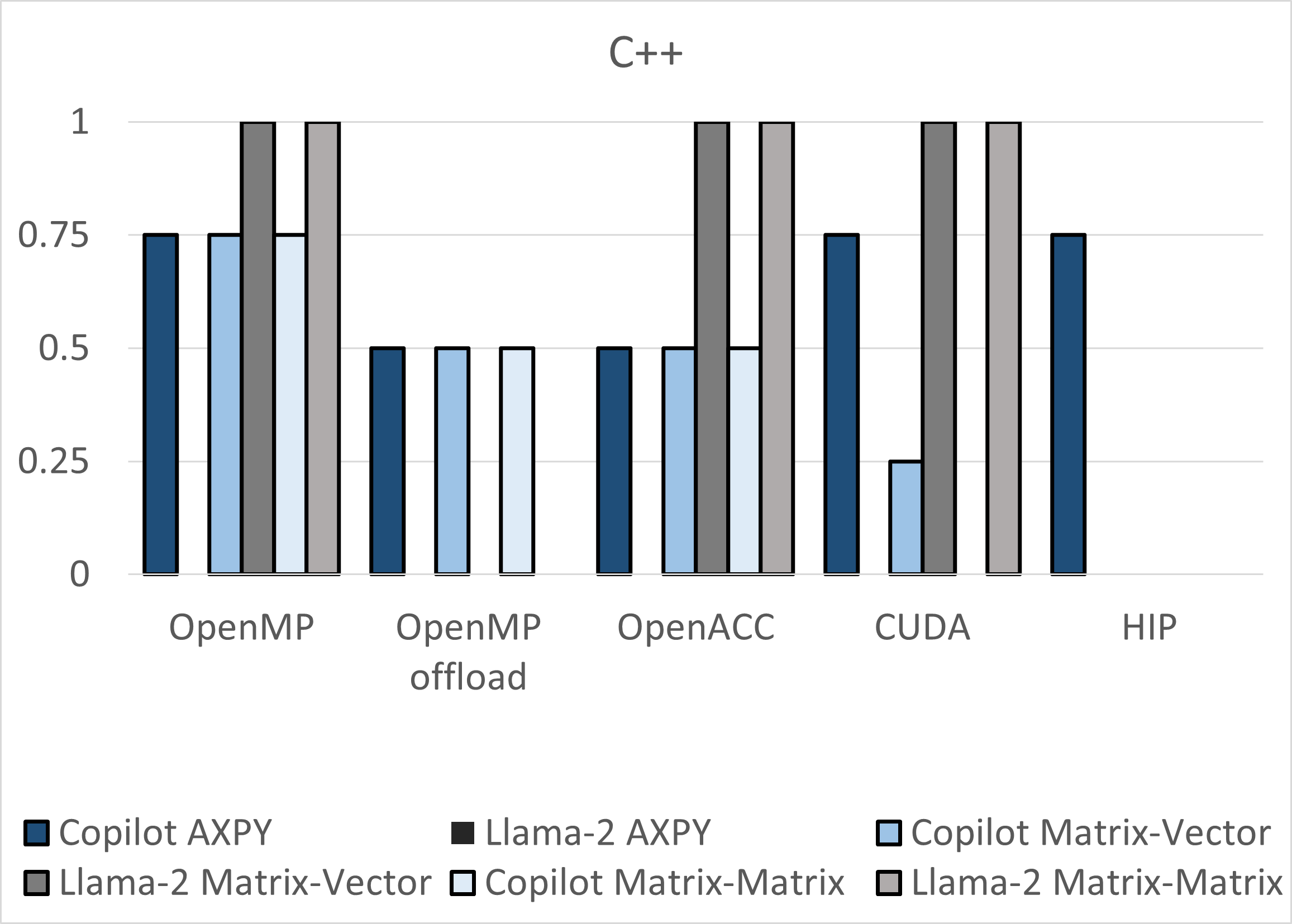} 
\includegraphics[width=0.48\textwidth]{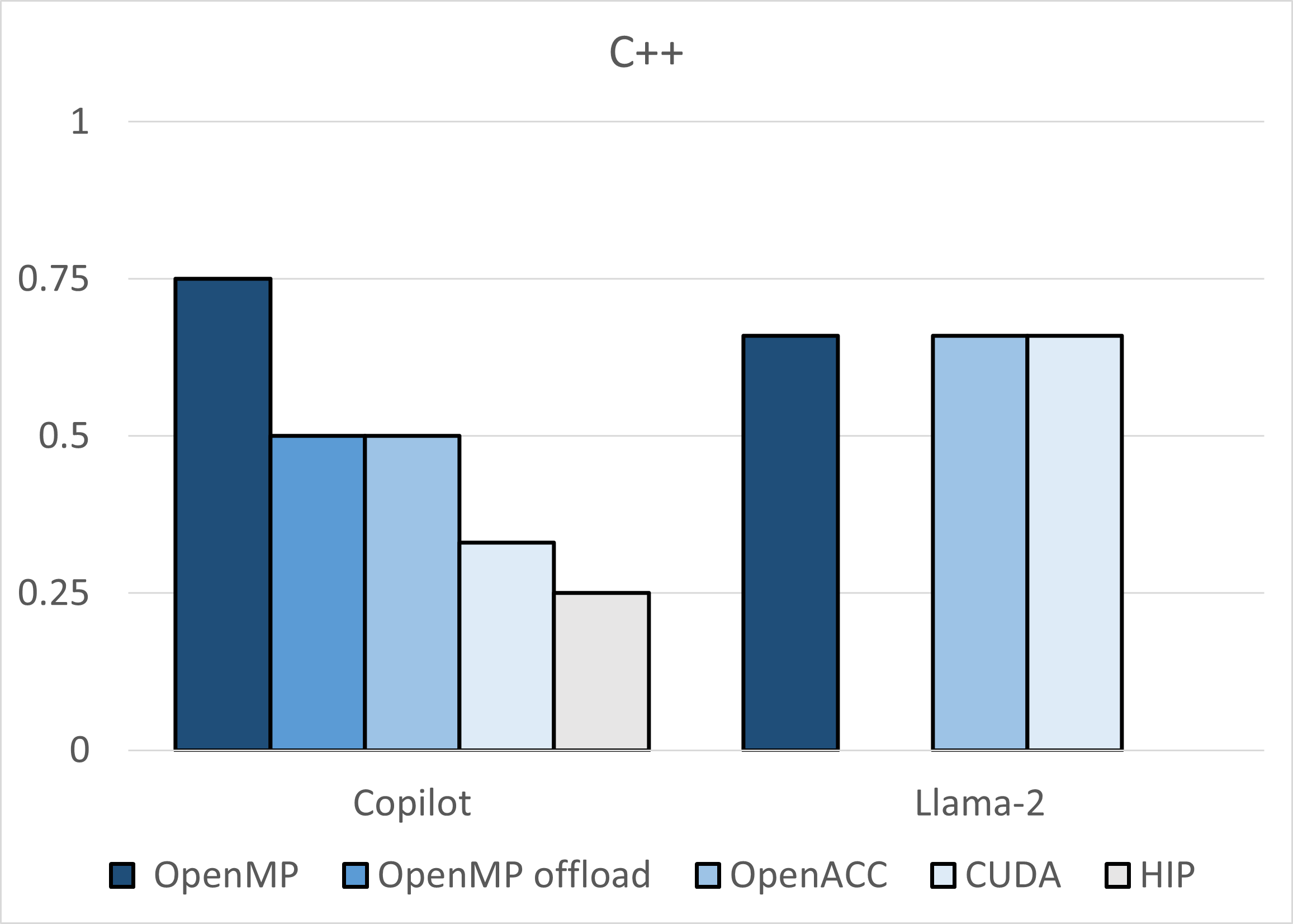} 
\caption{Results for C++ kernels (left) and programming models (right).}
\label{fig:c++}
\end{figure}

First, we focus on C++. Figure~\ref{fig:c++} illustrates the results (correctness) of the
C++ codes generated for OpenMP, OpenMP offload, OpenACC, CUDA, and HIP. 
As shown, Copilot can provide at least one correct code for most of the kernels and programming models, whereas Llama-2 
provided correct codes for OpenMP, OpenACC, and CUDA.
Although Llama-2 was unable to provide correct codes for OpenMP offload and HIP, the codes that it did correctly generate were 
higher quality (i.e., optimized) than the ones generated by Copilot. 

\begin{figure}[htb]
\centering
\includegraphics[width=0.48\textwidth]{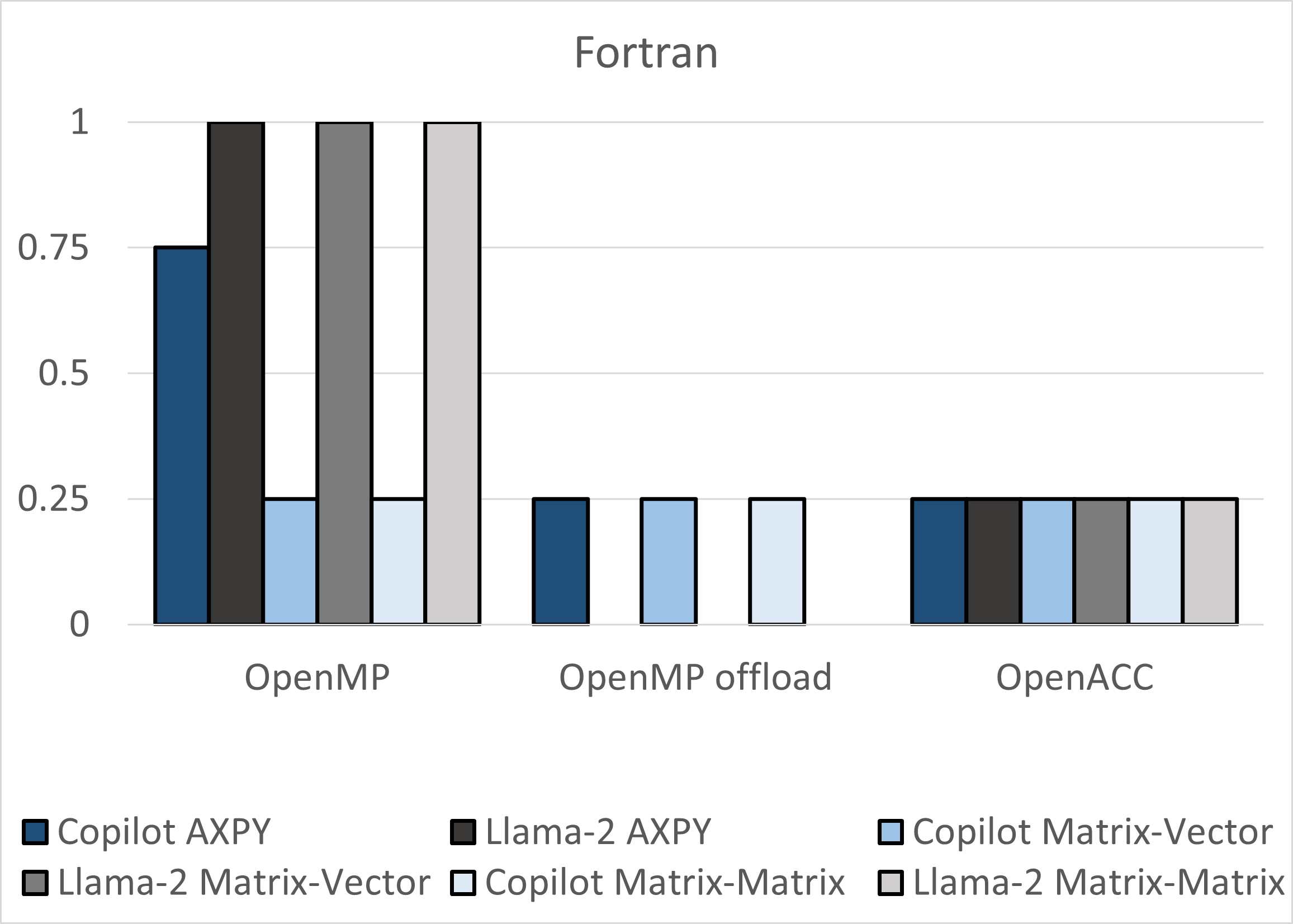} 
\includegraphics[width=0.48\textwidth]{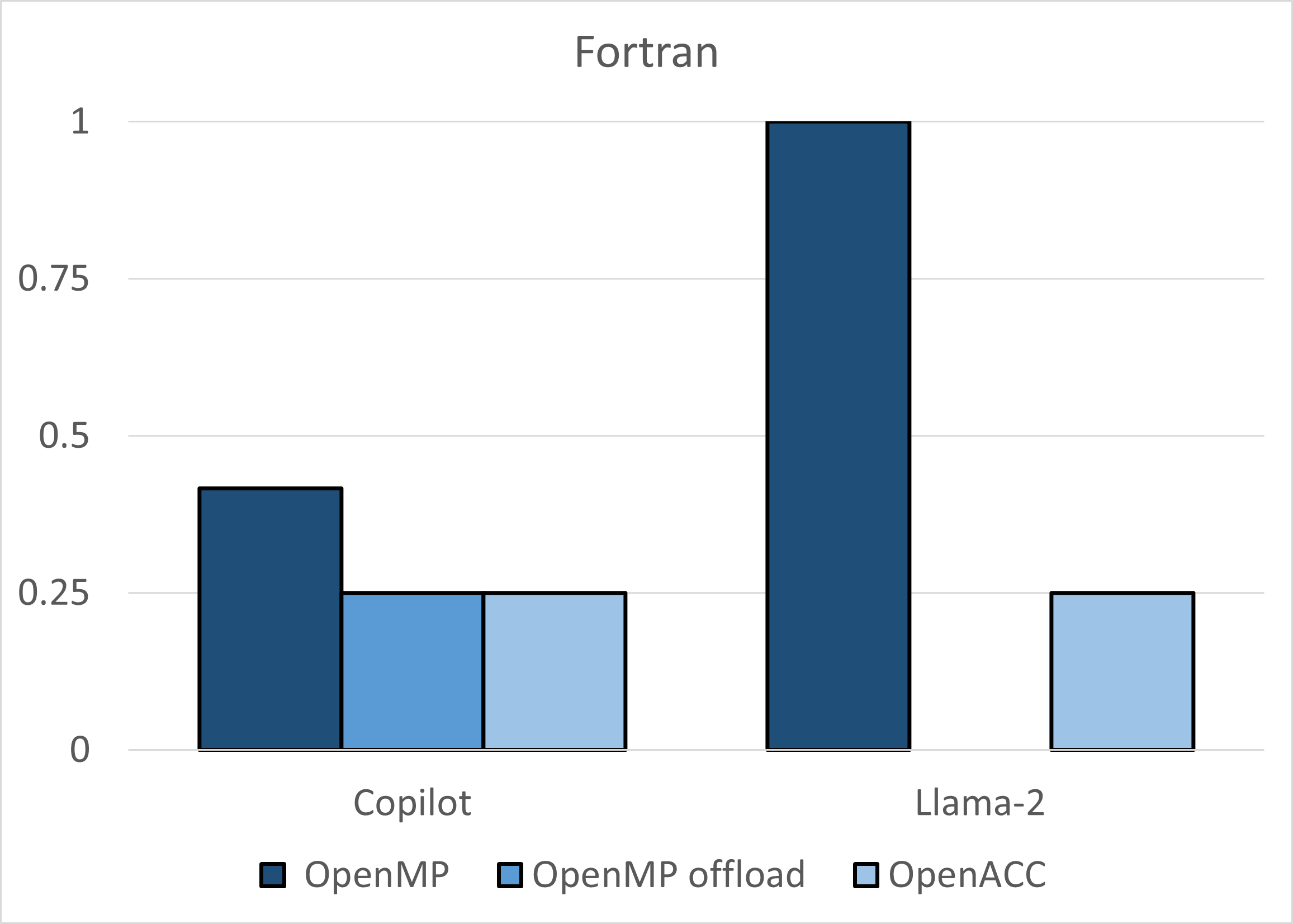}
\caption{Results for Fortran kernels (left) and programming models (right).}
\label{fig:fortran}
\end{figure}

For Fortran (Figure~\ref{fig:fortran}), we have a similar conclusion to that of the C++ study, with the exception of the AXPY kernel.
Here, we actually see that Llama-2 achieved much better performance for the AXPY kernel. Once again, however, Llama-2 provided very poor performance for OpenMP offload. 
Notably, Llama-2 generated high-quality OpenMP codes for all kernels.
Copilot still generated at least one correct code for all kernels and programming models and provided the same quality except for the AXPY-OpenMP test case.

\begin{figure}[htb]
\centering
\includegraphics[width=0.48\textwidth]{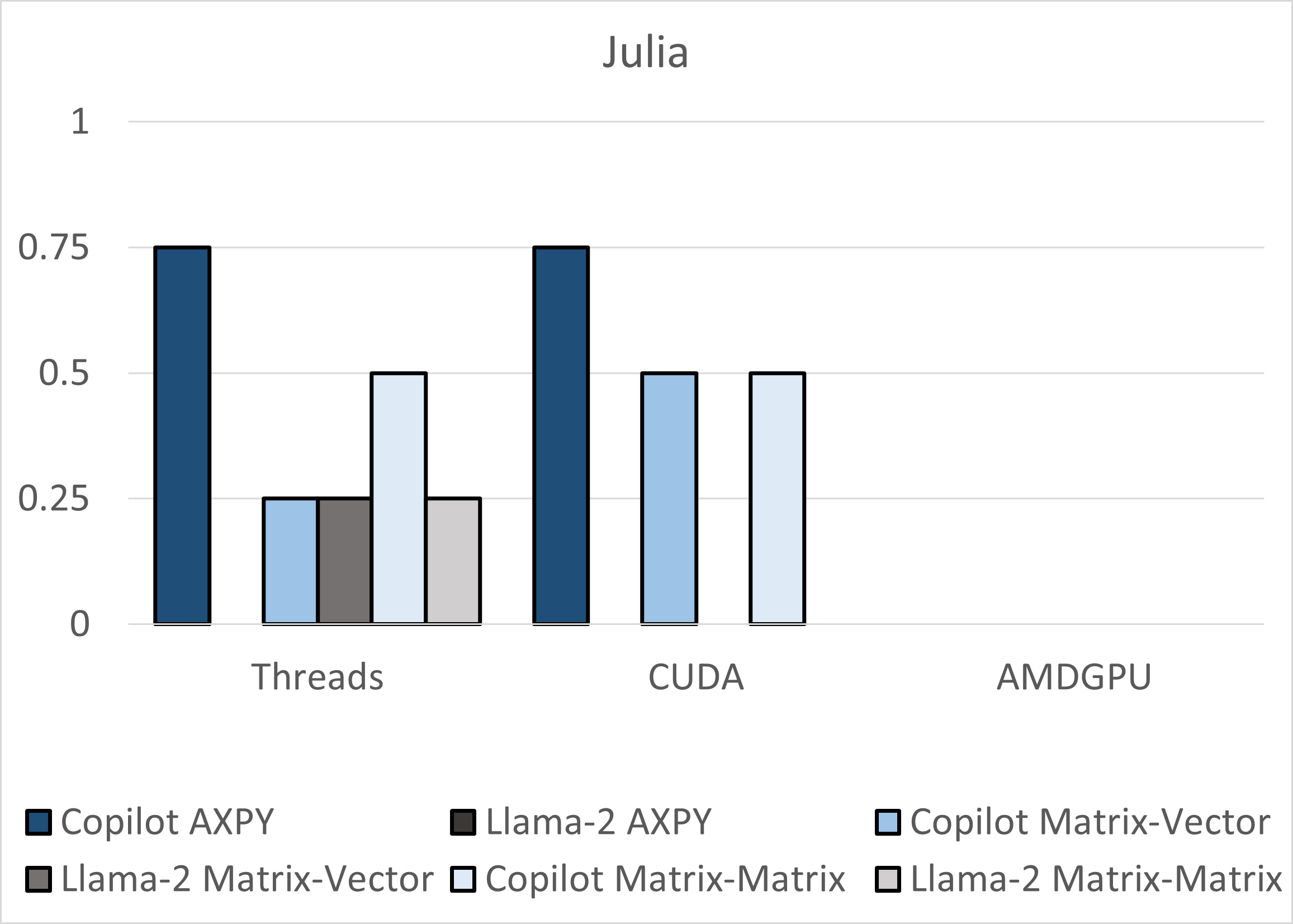} 
\includegraphics[width=0.48\textwidth]{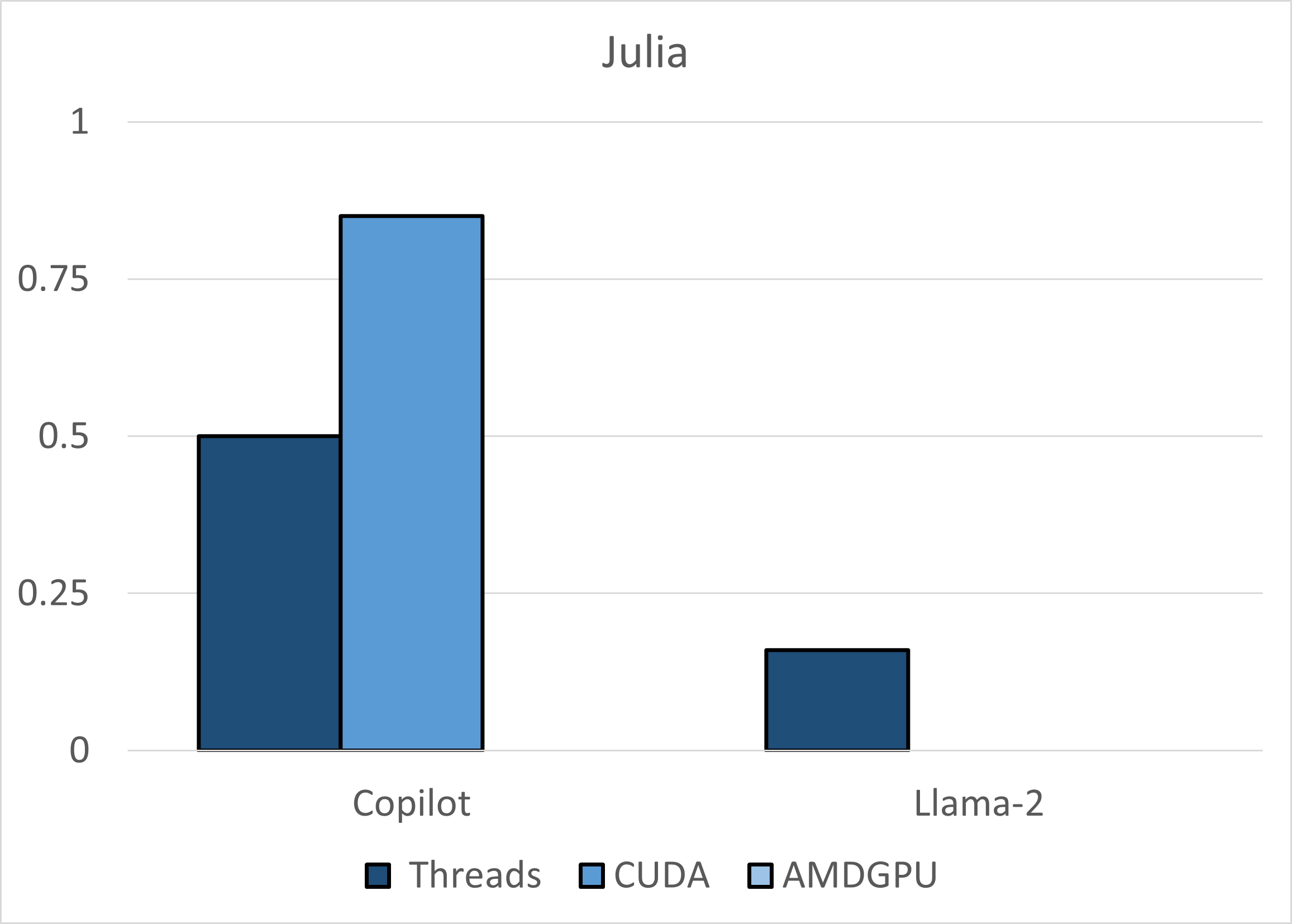}
\caption{Results for Julia kernels (left) and programming models (right).}
\label{fig:julia}
\end{figure}

For Julia, Llama-2 did not generate correct codes for any of the test cases  with the exception of the matrix-vector and matrix-matrix multiplications using the Base.Threads.jl Julia package. This case contained at least one correct code (Figure~\ref{fig:julia}). Unlike Llama-2, GitHub Copilot provided correct codes for all tests except for AMDGPU.jl, for which neither LLM was able to generate correct codes.

\begin{figure}[htb]
\centering
\includegraphics[width=0.48\textwidth]{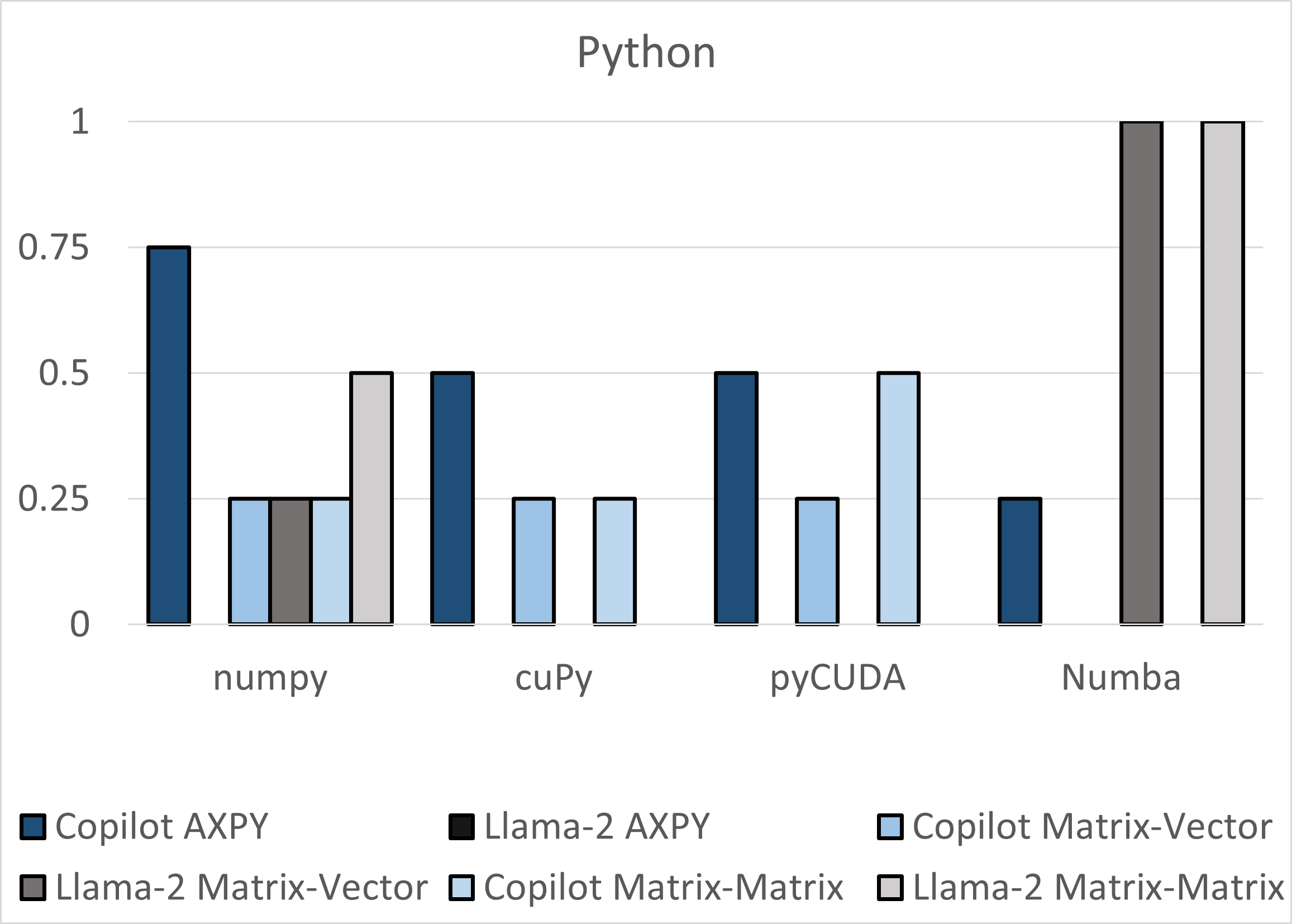} 
\includegraphics[width=0.48\textwidth]{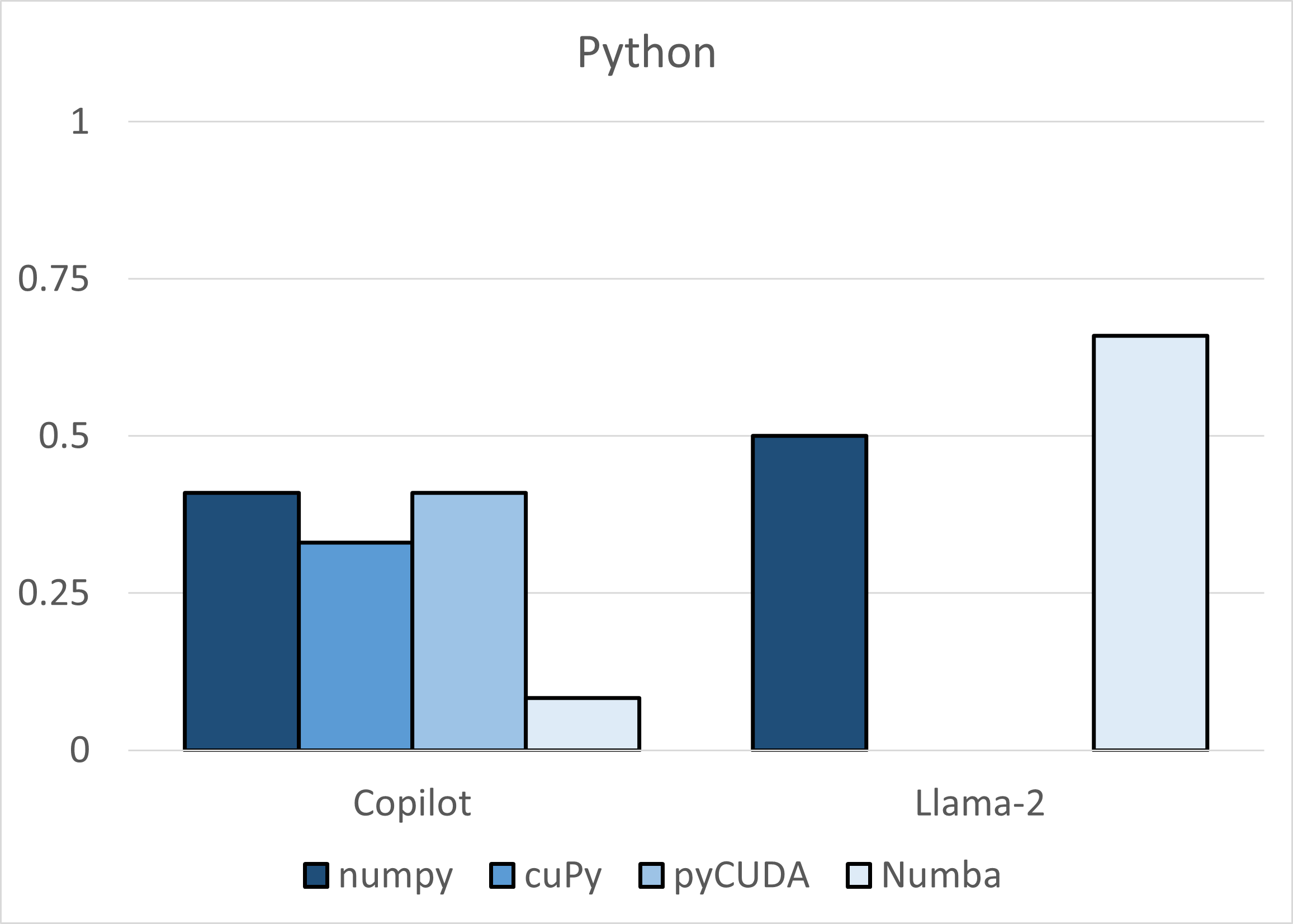}
\caption{Results for Python kernels (left) and programming models (right).}
\label{fig:python}
\end{figure}

Finally, Figure~\ref{fig:python} illustrates the results for the Python codes.
Copilot was able to generate at least one correct code for most of the test cases with the exception of the level-2 and level-3 BLAS kernels using Numba. Llama-2 provided the best results for these cases. Llama-2 also generated correct results for some numpy codes.

Overall, the main difference between the Copilot and Llama-2 LLMs is that, although Copilot can provide at least one correct code for most of the programming languages and models (albeit the generated codes are not optimized), Llama-2 is more aggressive in terms of optimizations, thereby providing well-optimized codes at the cost of generating incorrect codes in multiple cases. So, in general, Copilot generates codes that are more reliable but less optimized, and codes generated by Llama-2 are less reliable but more optimized.
\section{Conclusions}
\label{sec:Conclusions}
We evaluated the Llama-2 model as an HPC code generator for different programming languages (e.g., C\texttt{++}, Fortran, Julia, and Python) 
and models used for multicore CPUs (e.g., OpenMP, Base.Threads.jl), NVIDIA GPUs (e.g., CUDA, CUDA.jl, OpenACC, numpy, cuPy, pyCUDA, and Numba), and AMD GPUs (e.g., HIP and AMDGPU.jl).

Llama-2 can provide good-quality HPC codes for some of the previously mentioned solutions. When compared with GitHub Copilot, we realized that the Llama-2 model attempts to provide more optimized codes at the cost of not being as reliable as Copilot. In this study, Llama-2 was able to generate at least one correct code for 40\% (C++), 66\% (Fortran), 22\% (Julia), and 33\% (Python) of the test cases. GitHub Copilot provided at least one correct code in 80\% (C++), 100\% (Fortran), 66\% (Julia), and 83\% (Python) of the same test cases.

\section*{Acknowledgment}
This work is funded by Bluestone, an X-Stack project in the DOE Advanced Scientific Computing Office with program manager Hal Finkel.

%
%
%
\bibliographystyle{splncs04}
\bibliography{paper}
\end{document}